\documentclass[twoside,11pt]{article}
\usepackage{jair, theapa, rawfonts}

\ShortHeadings{GradSTL: Comprehensive STL for Neurosymbolic Reasoning \& Learning}
{Chevallier, Smola, Schmoetten, \& Fleuriot}
\firstpageno{1}

\usepackage{amsfonts}
\usepackage{amsmath}
\usepackage{amssymb}
\usepackage{amsthm}
\usepackage{listings}
\usepackage[normalem]{ulem}
\usepackage{mathtools}
\usepackage[most]{tcolorbox}
\usepackage{tikz}
\usetikzlibrary{matrix}
\usetikzlibrary{decorations.pathreplacing}
\usepackage{color,soul}
\usepackage{soul}
\usepackage{url}
\usepackage[utf8]{inputenc}
\usepackage{graphicx}
\usepackage{booktabs}
\usepackage{algorithmic}

\newcommand\Item[1][]{%
  \ifx\relax#1\relax  \item \else \item[#1] \fi
  \abovedisplayskip=0pt\abovedisplayshortskip=0pt~\vspace*{-\baselineskip}}

\theoremstyle{plain}

\newcommand{\ltlf}{LTL$_f$}

\tcbset{colback=gray!10, colframe=black!60, boxrule=0.5pt, arc=2pt, left=6pt, right=6pt, top=4pt, bottom=4pt}

\theoremstyle{plain}

\newtheorem*{theorem*}{Theorem}
\newtheorem{theorem}{Theorem}

\usepackage{enumitem}
\usepackage{regexpatch}

\title{GradSTL: Comprehensive Signal Temporal Logic for Neurosymbolic Reasoning and Learning}

\author{\name Mark Chevallier \email mchevall@ed.ac.uk \\
       \name Filip Smola \email f.smola@ed.ac.uk \\
       \name Richard Schmoetten \email richard.schmoetten@ed.ac.uk \\
       \name Jacques D. Fleuriot \email jdf@ed.ac.uk}

\begin{document}

\maketitle

\begin{abstract}
We present GradSTL, the first fully comprehensive implementation of signal temporal logic (STL) suitable for integration with neurosymbolic learning. In particular, GradSTL can successfully evaluate any STL constraint over any signal, regardless of how it is sampled. Our formally verified approach specifies smooth STL semantics over tensors, with formal proofs of soundness and of correctness of its derivative function. Our implementation is generated automatically from this formalisation, without manual coding, guaranteeing correctness by construction. We show via a case study that using our implementation, a neurosymbolic process learns to satisfy a pre-specified STL constraint. Our approach offers a highly rigorous foundation for integrating signal temporal logic and learning by gradient descent.
\end{abstract}

\section{Introduction}
\label{sIntro}

Signal temporal logic (STL) \cite{maler2004monitoring} is a formal language suitable for expressing a wide variety of temporal constraints with relevance to control methods \cite{raman2015reactive}, robotics \cite{abele2016time}, and more recently, machine learning \cite{kapoor2025stlcgmaskingapproachdifferentiable,leung2023backpropagation}. These constraints can be used to describe safety conditions, for example in autonomous vehicles \cite{arechiga2019specifying}. In neurosymbolic processes, STL can be used to define interpretable objectives or constraints over learning. However, previous implementations of differentiable STL support either only a part of the language or rely on assumptions about the signal it is applied to (for example, assuming uniform time sampling). This fails to achieve the full potential of STL, limiting its applicability.

We introduce GradSTL, the first comprehensive and formally verified implementation of a differentiable semantics of STL expressed over tensors, applicable to any finite signal and suitable for immediate use in neurosymbolic learning. By ``comprehensive'', we mean a full implementation of all aspects of the STL language, syntax and full semantics, usable with any signal. Using the Isabelle theorem prover \cite{nipkow2002isabelle}, we formalise the STL language and its standard semantics, its robustness and the derivative of its robustness, all in an algorithmic form which recurses over arbitrary signals.

We formally prove both the soundness of the robustness function with respect to STL's standard semantics, and the correctness of the derivative function. GradSTL's specification is comprehensive, allowing for nested temporal constraints, the Until constraint and making no additional assumptions about the signals it is defined over. We have found no previous implementation of STL for learning via gradient descent that has achieved all of these properties (see Section \ref{ssBackRelated}). To do this, we use a novel \emph{adaptive temporal window} technique to change temporal constraints as they recurse down a signal temporally (see Section \ref{ssFormSTLsemantics}). 

From this specification, we automatically generate executable OCaml code, avoiding any unverified manual implementation. This integrates directly with PyTorch-based learning processes, allowing the STL robustness value to be used as a loss function in (for example) a neural network. We demonstrate experimentally that we can then learn behaviours satisfying arbitrary STL constraints.

Our contribution is both theoretical and practical: a sound and comprehensive STL semantics for recursing over irregularly sampled signals; a formally verified, differentiable robustness function; a verified-to-executable implementation pipeline; and empirical validation in a modern machine learning framework. By eliminating ad hoc and unverifiable logic handling, our work offers a rigorous and trustworthy foundation for constrained neural learning and formal specification-guided AI.

This work is an extension of previous work integrating formalised linear temporal logic over finite traces (\ltlf{}) into neurosymbolic learning \cite{chevallier2025formallyverifiedneurosymbolictrajectory}. It expands on this previous work substantially, with a novel approach to working with temporal constraints in STL, as well as expanding the applicability of STL to learning as discussed above. The most important innovations over the previous work are a method of evaluating differentiable functions as atomic constraints, and a temporal recursion method required for STL to work well over irregularly sampled time periods (discussed further in Section \ref{ssFormSTLsemantics}). The current work also demonstrates how the same pipeline of formal verification methods and code generation can be successfully used with a substantially more complicated logic.

\subsection{Organisation of the paper}

In the next section, we briefly introduce STL and discuss the background to our work. Then, in Section \ref{sFormalisation}, we present our formal specification of a recursive algorithm to evaluate STL and statements of its formally proven properties. We also detail how we were able to overcome previous limitations using the adaptive temporal window technique. We also briefly describe the automatic code generation for our algorithms and how to integrate them into a learning process. In Section \ref{sExperiments}, we consider a case study showing the comprehensiveness of our implementation, concentrating specifically on irregular sampling.~Lastly, in Section \ref{sConclusion}, we discuss the significance of our work.

\section{Background}
\label{sBack}

In this section, we introduce signal temporal logic before going on to examine previous work integrating it with neurosymbolic processes.

\subsection{Signal temporal logic: a brief overview}
\label{ssBackSTL}

STL is a formal language intended to make a statement $\rho$ about some continuous function of time $C$ \cite{maler2004monitoring}. This function $C$ can model anything that changes over time: performance in an engine, the location of a vehicle, or health statistics of a patient, for example. The state at time $t$, $C(t)$, is typically a vector in $\mathbb{R}^m$ of state variables $(v_0, v_1, \dots , v_{m-1})$ representing measures of interest -- from our examples, perhaps these measures are the heat in the fuel chamber, the $x$-coordinate of a path, or the systolic blood pressure of a patient. This vector is called the \emph{sample} at time $t$ of $C$. 

In this paper, we examine two ways of assigning meaning to an STL constraint: the \emph{standard} boolean semantics and the \emph{robustness} (see below). The standard semantics for STL evaluates a statement $\rho$ as being either satisfied (\emph{true}) or unsatisfied (\emph{false}). Such STL statements are often called \emph{constraints} as they describe conditions that might be breached or respected by $C$. This evaluation takes place in the context of a temporal \emph{signal} which summarises $C$.

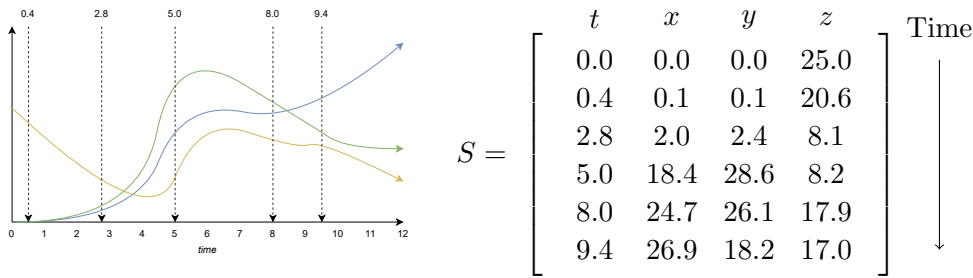
\begin{figure}[hbtp]
    \centering
    \begin{minipage}[b]{0.4\textwidth}
    \includegraphics[width=1.0\textwidth]{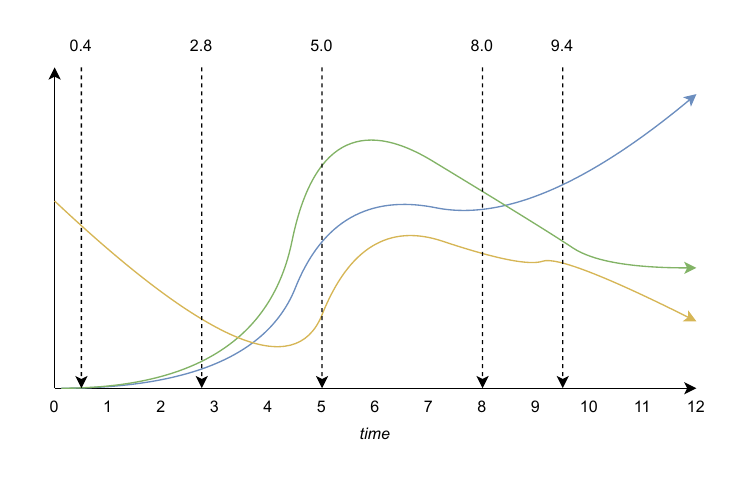}
    \end{minipage}
    \begin{minipage}[b]{0.59\textwidth}
    \begin{tikzpicture}
        \matrix[
            matrix of math nodes,
            nodes={text width=2em, align=center},
            left delimiter={[},right delimiter={]},
        ] (m) {
            0.0 & 0.0 & 0.0 & 25.0 \\
            0.4 & 0.1 & 0.1 & 20.6 \\
            2.8 & 2.0 & 2.4 & 8.1 \\
            5.0 & 18.4 & 28.6 & 8.2 \\
            8.0 & 24.7 & 26.1 & 17.9 \\
            9.4 & 26.9 & 18.2 & 17.0 \\
        };
        \pgfmathsetmacro{\hspacing}{28}
        \pgfmathsetmacro{\vspacing}{8}
        \draw[->] ([xshift=\hspacing pt]m-1-4.east) -- ([xshift=\hspacing pt]m-6-4.east)
            node[at start, anchor=north, yshift=20pt]{Time};
        \draw
            ([yshift=\vspacing pt]m-1-1.north) node{$t$}
            ([yshift=\vspacing pt]m-1-2.north) node{$x$}
            ([yshift=\vspacing pt]m-1-3.north) node{$y$}
            ([yshift=\vspacing pt]m-1-4.north) node{$z$};
        \draw
            ([xshift=-28pt,yshift=\vspacing pt]m-4-1.west) node{$S=$};
    \end{tikzpicture}
    \end{minipage}
    \caption{Example of a signal sampling three state variables $x$ (in blue), $y$ (in green), and $z$ (in yellow), at times 0.0, 0.4, 2.8, 5.0, 8.0 and 9.4. An illustration of sampling is on the left, and the signal in matrix form is on the right.}
    \label{fig:siggraph}
\end{figure}

A signal $S_{C,T}=(C(t_0),C(t_1),\dots,C(t_{n-1}))$ is a vector of samples of $C$ where $T$ is a vector of \emph{time indices} in $\mathbb{R}^n$, $(t_0, t_1,\dots, t_{n-1})$, ordered so that for all $i<j$ we have $t_i < t_j$. These time indices $t_i$ hold the time when each sample was taken, and can be arbitrarily close together or far apart. We typically drop the subscripts $C,T$ where there is no ambiguity. As $S$ is a vector of vectors, we can represent it as a matrix (a two dimensional tensor). Figure \ref{fig:siggraph} shows how a signal samples an underlying function of three state variables.

The precise time indices used for the samples can make a particular constraint true or false, even over the same underlying function $C$.  Thus, effective sampling is critical to ensure that the signal reflects $C$ sufficiently well for the constraint to be useful.

In the remainder of this paper, we will use the notation $t\leftarrow S$ to express that $t$ is a time at which signal $S$ is sampled, in other words, that $t$ is a component of the time vector $T$ used by $S$. We define $\lvert S \rvert$ as the total number of samples in the signal $S$.

We define $S_n=C(t_n)$ if $t_n \leftarrow S$. We denote the standard semantic evaluation of a constraint $\rho$ at time $t_n\leftarrow S$ over signal $S$ by $E_S(\rho, n)$ and assume that $n < \lvert S \rvert$, otherwise $E_S$ is undefined.

As well as the standard semantics, STL also can be evaluated via a \emph{robustness} semantics, which is also considered over a signal. This returns a real value whose magnitude corresponds, roughly speaking, to how much the signal would need to be changed for the constraint's truth value to change from true to false (or vice versa). If robustness is negative for some constraint $\rho$, then $\rho$ is false. If it is positive, then $\rho$ is true. If robustness is exactly $0$ for $\rho$, we do not know if it is true or false -- it is on the cusp and might be either. We denote the robustness of a constraint $\rho$ at time $t_n\leftarrow S$ by $R_S(\rho, n)$ and, as for $E_S$, we assume $n < \lvert S \rvert$.

Syntactically, STL constraints are represented by logical formulae. The smallest constraint possible is an atomic one which, when evaluated at sample $n$, compares $f(S_n)$ to some constant $c$, where $f$ is some differentiable real-valued function. We denote these constraints by $\mu_{f,c}$ whose semantics depends on the function $f$ and constant $c$. For example, if we are monitoring the location of a robot, and we can extract its $x$ and $y$ coordinates from $S_n$, we can calculate its distance from the origin using $f(x,y) = \sqrt{x^2+y^2}$ to evaluate if it has travelled past some boundary distance. The semantics of an atomic constraint is then defined as follows:$$
    E_S(\mu_{f,c}, n) \Coloneq f(S_n) > c$$

\noindent Building on this, the STL formulae used to capture the more general constraints are:$$
\rho \Coloneq \mu_{f,c} \mid \neg \rho \mid \rho_1 \wedge \rho_2 \mid \square_{[x,y]} \rho \mid \Diamond_{[x,y]} \rho \mid \rho_1 \mathbin{\mathcal{U}_{[x,y]}} \rho_2$$

\noindent The first three are understood as per the syntax and semantics of propositional logic, with $\mu_{f,c}$ being the atomic constraint (proposition), and $\neg \rho$ and $\rho_1 \mathbin{\wedge} \rho_2$ representing negation and conjunction, respectively. Disjunction can be represented by $\neg (\neg\rho_1 \wedge \neg\rho_2)$.

The remaining formulae introduce various temporal constraints. Each of these has a subscript $[x,y]$, with $0\leq x \leq y$, which shows that when the constraint is evaluated at a sample with time index $t$, it applies at times sampled between $t+x$ and $t+y$. These can be understood as follows:

\begin{description}
	\item[Always] {
            \begin{align*}
                E_S(\square_{[x,y]} \rho, n) \Coloneq \forall t_i\leftarrow S.\; t_n+x\leq t_i \leq t_n+y \implies E_S(\rho, i)
            \end{align*}
        }
        In other words, when evaluated at a sampled time $t_n$, $\square_{[x,y]} \rho$ is true if and only if $\rho$ is true at all sampled times between and including $t_n+x$ and $t_n+y$.\\
        
	\item[Eventually] { \begin{align*}E_S(\Diamond_{[x,y]} \rho, n) \Coloneq \exists t_i\leftarrow S.\; t_n+x\leq t_i \leq t_n+y \wedge E_S(\rho, i)\end{align*}  When evaluated at a sampled time $t_n$, $\Diamond_{[x,y]} \rho$ is true if and only if $\rho$ is true at some sampled time between and including $t_n+x$ and $t_n+y$.}\\
            
    \item[Until] { \begin{align*}
    E_S(\rho_1\mathbin{\mathcal{U}_{[x,y]}}\rho_2, n) \Coloneq\ \exists t_i\leftarrow S.\; & t_n+x\leq t_i \leq t_n+y \wedge E_S(\rho_2, i) \; \wedge \\ & \bigl(\forall t_j\leftarrow S.\; t_n+x\leq t_j \leq t_i \rightarrow \;E_S(\rho_1, j)\bigr)
    \end{align*}
    When evaluated at a sampled time $t_n$, $\rho_1\mathbin{\mathcal{U}_{[x,y]}}\rho_2$ is true if and only if $\rho_2$ is true at some time between $t_n+x$ and $t_n+y$, and $\rho_1$ is true for all sampled times between and including $t_n+x$ and whenever $\rho_2$ is first true.}
\end{description}

\noindent Our treatment of STL is not minimal -- we could define the language without expressly including the $\square_{[x,y]} \rho$ and $\Diamond_{[x,y]} \rho$ as primitive constraints, instead using $\neg(\top \mathbin{\mathcal{U}_{[x,y]}}\neg\rho)$ and $\top\mathbin{\mathcal{U}_{[x,y]}}\rho$, respectively (where $\top$ is True which can be defined to be any tautology). We provide these constraints and their semantics explicitly to allow more efficient code generation (see Section \ref{ssFormCodeGen}).

Robustness is calculated using the $R_S(\rho, n)$ function, which returns a real value that, as noted above, corresponds to how well the constraint is satisfied over the signal. Its definition  over our STL syntax is as follows:

\begin{itemize}
    \item $R_S(\mu_{f,c}, n) = f(S_n) - c$
    \item $R_S(\neg\rho, n) = -R_S(\rho, n)$
    \item $R_S(\rho_1\wedge\rho_2, n) = \min \{R_S(\rho_1, n),\;R_S(\rho_2, n)\}$
    \item $R_S(\square_{[x,y]} \rho, n) = \min \{R_S(\rho,i): i < \lvert S \rvert \wedge t_n+x\leq t_i \leq t_n+y\}$
    \item $R_S(\Diamond_{[x,y]} \rho, n) = \max \{R_S(\rho,i): i < \lvert S \rvert \wedge t_n+x\leq t_i \leq t_n+y\}$
    \Item \begin{flalign*}
    R_S(\rho_1\mathcal{U}_{[x,y]} \rho_2, n) = & \\
    \max \Bigl\{&\min\bigl\{R_S(\rho_2,i),\min \{R_S(\rho_1,j): j < \lvert S \rvert \wedge t_n+x\leq t_j \leq t_n+t_i\}\bigr\}\\&:i < \lvert S \rvert \wedge t_n+x\leq t_i \leq t_n+y\Bigr\}
    \end{flalign*}
\end{itemize}

\noindent Examining these sub-definitions, one can prove by induction, through both the constraint structure and the temporal dimension, that whenever $R_S(\rho, n)>0$, $E_S(\rho, n)$ is true, and that whenever $R_S(\rho, n)<0$, $E_S(\rho, n)$ is false \cite{maler2004monitoring}. When $R_S(\rho, n)=0$, it also holds that $R_S(\neg\rho, n)=0$, so this tells us nothing about $E_S(\rho, n)$. We call this $R_S$ \emph{sound} with respect to $E_S$.
Soundness is one of many useful (in this case, essential) properties a robustness can satisfy \cite{flinkow2024comparing}. In Section~\ref{ssFormSTLsmooth}, we give an algorithmic specification of a semantics similar to the above, but which is differentiable and recurses efficiently.

The definition of $R_S$ can be derived from that of $E_S$ following some simple principles. The $R_S$ rule for the atomic constraint is fundamental and clearly follows from the equivalent rule for $E_S$ (likewise for negation).
For the others, we take minimums of sets of $R_S$ values to represent the universal quantifier over (resp.\ conjunctions between) their corresponding $E_S$ truth values, and maximums to represent the existential quantifier (resp.\ disjunction).
If we conjoin two values, and one is false -- implying that its $R_S$ value is non-positive -- the result will also be non-positive (similar reasoning for disjunctions).

It is important to note that these definitions given for $E_S$ and $R_S$ cannot easily be translated to an efficient, recursive algorithm that computes its value over a signal. GradSTL uses provably equivalent definitions of these functions that give us recursive algorithms that compute over arbitrary signals efficiently. We discuss the details in Sections \ref{ssFormSTLsemantics} and \ref{ssFormSTLsmooth}.

\subsection{Previous work on neurosymbolic integration of STL}
\label{ssBackRelated}

There are several previous approaches to integrating STL with neurosymbolic learning. Our work is aimed at using an arbitrary STL constraint to learn signal outputs which satisfy that constraint. Not all neurosymbolic work using STL has the same goal, and, to our knowledge, no previous neurosymbolic approach provides a comprehensive implementation of STL against arbitrary signals. We next review previous work in this area, assessing each approach in terms of its goal and how comprehensively it implements STL.

Leung et al.~worked on predicting a STL constraint (given a fixed structure) would best satisfy a given signal \cite{leung2019backpropagation}. The goal of this work was to learn the numeric parameters for the fixed STL constraint using neural methods and a parametric STL approach \cite{asarin2012parametric}. Computation graphs implemented through the PyTorch library were fundamental to this work, which assumed that any signal was uniformly sampled.

The use of computation graphs is also a hallmark of later work by the same authors on STLCG \cite{leung2023backpropagation}, and work on its successor STLCG++ \cite{kapoor2025stlcgmaskingapproachdifferentiable}. Of the work discussed in this section, these two methods are the closest to our own -- aimed at providing a way to use satisfaction of an STL constraint in learning some temporal output. Backpropagation using these computation graphs is key to how they enable learning. However, STLCG and its successor both assume uniform time sampling across the signal, which means they can only be applied to a proper subset of signals. In contrast, as discussed in Section~\ref{ssFormSTLsemantics}, our own work can be applied to signals with any sampling.

Yan et al.~worked on STONE, a method aimed at learning weighted STL subconstraints that efficiently classify time series \cite{yan2021stone}. Rather than using an STL constraint to help learn a temporal output, this work uses a temporal input (a time series) to learn the weights for a STL constraint's subconstraints, to help interpret and classify subsequent time series. Each subconstraint is represented as an individual neuron within a network, integrating the constraint with the neural process. However, the STL implemented by STONE is only a fragment of STL -- it omits the important Until constraint, and requires its initial structure to be configured ahead of time; only the weights are learned.

X.~Li et al.~have enhanced the work on STL constraint classification, but their approach retains the limitation of only considering a fragment of STL, ignoring the Until constraint \cite{li2024tlinet}. This inability to account for Until is shared by the work on differentiable logic layers by D.~Li et al.~\cite{li2021differentiable}. Again, aimed at learning an STL constraint using temporal inputs, this uses a layer of the neural network to represent the STL constraint itself, adjusting the layer's weights to represent how the constraint might function. As these methods use the structure of the neural network to learn the constraint, what they learn is limited by said structure but also the size of the neural network. They cannot learn general constraints.

Liu et al.~use earlier work on BarrierNets \cite{xiao2021barriernet} to guarantee that the output of a given process will satisfy an STL constraint \cite{liu2023learning}. These approaches have a similar goal to our own work, but the use of BarrierNets provides a guarantee that a given STL constraint will be satisfied by modifying the neural network's output. Again, this uses a fragment of STL ignoring the Until constraint, and also enforces restrictions on the domain it can work with: those with safety methods involving reaching or avoiding circular regions.

As can be seen, no previous work in this area has managed to deal with a fully comprehensive treatment of STL. Either the language is limited (excluding the Until constraint or lacking nested temporal constraints), or its applicability relies on significant assumptions (working only with uniformly sampled signals). It is precisely these limitations GradSTL overcomes.

Outside of work applicable to neurosymbolic learning, other STL monitors like Breach \cite{donze2010breach} and S-TaLiRo \cite{annpureddy2011s} can work with irregularly sampled signals, but unlike GradSTL, they find a non-differentiable robustness and are only used for falsification. They are also coded by hand, leading to the potential for error.

\section{Formal Semantics for GradSTL}
\label{sFormalisation}

We now discuss the details of our algorithmic specification of STL and its properties in two sections: first we discuss our work specifying an algorithm for STL semantics in the Isabelle theorem prover, presenting a simple algorithm to compute its boolean semantics and showing how, from this, we can derive a function for its robustness that is differentiable. Secondly, we briefly review code generation and how our formalisation feeds into it.

\subsection{Recursive algorithm for the standard semantics of STL}
\label{ssFormSTLsemantics}

\begin{tcolorbox}[title=Notation]
\renewcommand{\arraystretch}{1.1}
\begin{tabular}{@{}ll@{}}
$S_n$ & Sample $n$ in signal $S$, expressed as a vector of state variables \\
$\lvert S \rvert$ & The total number of temporal samples in $S$ \\
$t \leftarrow S$ & Time $t$, where $S$ contains a sample at time $t$ \\ 
$\Delta t_n$ & $t_{\,n+1} - t_n$ (time gap between sample $n$ and $n+1$) \\
\end{tabular}
\end{tcolorbox}

We now examine how to algorithmically evaluate the semantics of an STL constraint over a signal. We specify the function $E^*_S(\rho, n)$ for this purpose, where $S$ is a signal, $\rho$ is the constraint being evaluated, and $n$ a natural number indexing the signal's samples. This algorithm is structured such that $E^*_S(\rho, n) = E_S(\rho, n)$. The evaluation proceeds by recursion through the constraint $\rho$. The propositional component of the constraint proceeds exactly as in Section \ref{ssBackSTL}:

\begin{itemize}
    \item $E^*_S(\mu_{f,c},n) \Coloneq f(S_n) > c$
    \item $E^*_S(\neg\rho,n) \Coloneq \neg\bigl(E^*_S(\rho,n)\bigr)$
    \item $E^*_S(\rho_1 \wedge \rho_2,n) \Coloneq E^*_S(\rho_1,n) \wedge E^*_S(\rho_2,n)$
\end{itemize}

\noindent The temporal constraints are more complicated. In addition to recursion through the constraint itself, $E^*_S$ may recurse down the signal's temporal dimension to evaluate at a later time. The key to correctly evaluating temporal constraints is the \emph{adaptive temporal window}, which changes as temporal recursion takes place. After each temporal position $n$, if $n<\lvert S \rvert -1$, we subtract $\Delta t_n$ from the temporal window, where $\Delta t_n = t_{n+1}-t_n$. This technique allows for nesting of temporal constraints to any depth, and the full expressiveness of STL over arbitrary signals. This overcomes the limitations of previous attempts that did not use this method at applying STL to neurosymbolic learning.

We illustrate this using $E^*_S(\Diamond_{[x,y]} \rho, n)$. Informally, this is true if and only if:

\begin{quote}
Case 1: $x \leq 0 \leq y \wedge E^*_S(\rho, n)$ is true. The first part of this case ($x \leq 0 \leq y$) checks if the temporal window contains $0$. If so, it means that if the second condition ($E^*_S(\rho, n)$) is true, we can confirm that the Eventually constraint has been satisfied (as defined in Section \ref{ssBackSTL}: $\exists t_i\leftarrow S.\; t_n+x\leq t_i \leq t_n+y \wedge E_S(\rho, i)$). If and only if this case is not true, then we check Case 2.

Case 2: $E^*_S(\Diamond_{[x-\Delta t_n,y-\Delta t_n]}\rho,n+1)$, is true. In other words, we re-evaluate at the next temporal position. The adaptive temporal window for the constraint changes, subtracting how much time difference there is between positions $n$ and $n+1$ ($\Delta t_n$). If this changes the temporal window so that it contains $0$, we know that the time index for the current position is within the temporal window, so Case 1 above might become true.
\end{quote}

\noindent We also perform certain boundary checks as we recurse, ensuring that $n<\lvert S \rvert$, i.e.~that we do not pass the final temporal position in the signal, leaving the function undefined if we breach this boundary check. We also terminate temporal recursion if we would otherwise exit the temporal window (as once $y<0$, we know that there are no more samples that could satisfy the constraint).

For example, consider an STL constraint of $\Diamond_{[5,10]} \mu_{v,20}$. This is true at time index $t$ in signal $S$ if and only if $\exists t'\leftarrow S.\; t+5 \leq t' \leq t+10$ where $v>20$ is true. The adaptive temporal window begins as $[5,10]$, but changes as it recurses to capture when it is in the correct position \emph{relative to the point of its first evaluation}. This is illustrated in Figure \ref{fig:temprec}.

\begin{figure}[hbtp]
    \centering
    \begin{tikzpicture}
        \matrix[
            matrix of math nodes,
            nodes={text width=2em, align=center},
            left delimiter={[},right delimiter={]},
        ] (m) {
            0.0 & 1.6 \\
            2.3 & 1.9 \\
            3.9 & 12.0 \\
            7.7 & 15.3 \\
            9.1 & 14.2 \\
            11.4 & 28.2 \\
        };
        \pgfmathsetmacro{\hspacing}{28}
        \pgfmathsetmacro{\vspacing}{8}
        \draw ([xshift=24pt]m-1-2.east) node{$2.3$}
            ([xshift=24pt]m-2-2.east) node{$1.6$}
            ([xshift=24pt]m-3-2.east) node{$3.8$}
            ([xshift=24pt]m-4-2.east) node{$1.4$}
            ([xshift=24pt]m-5-2.east) node{$2.3$}
            ([xshift=24pt]m-6-2.east) node{$\dots$};
        \draw
            ([yshift=\vspacing pt]m-1-1.north) node{$t$}
            ([yshift=\vspacing pt]m-1-2.north) node{$v$}
            ([xshift=38pt,yshift=\vspacing pt]m-1-2.north) node{$\Delta t$};
        \draw 
            ([xshift=-10pt]m-1-1.west) node[anchor=east, align = right]{$E^*_S(\Diamond_{[5,10]}\mu_{v,20},0)\longrightarrow$}
            ([xshift=-10pt]m-2-1.west) node[anchor=east, align = right]{$E^*_S(\Diamond_{[2.7,7.7]}\mu_{v,20},1)\longrightarrow$}
            ([xshift=-10pt]m-3-1.west) node[anchor=east, align = right]{$E^*_S(\Diamond_{[1.1,6.1]}\mu_{v,20},2)\longrightarrow$}
            ([xshift=-10pt]m-4-1.west) node[anchor=east, align = right]{$E^*_S(\Diamond_{[-2.7,2.3]}\mu_{v,20},3)\longrightarrow$}
            ([xshift=-10pt]m-5-1.west) node[anchor=east, align = right]{$E^*_S(\Diamond_{[-4.1,0.9]}\mu_{v,20},4)\longrightarrow$};
    \end{tikzpicture}
    \caption{How $E^*_S(\Diamond_{[5,10]}\mu_{v,20},0)$ changes as it temporally recurses over $S$. At each recursion, the value $\Delta t$ is subtracted from the adaptive temporal window. When the temporal window contains $0$, $\mu_{v,20}$ is evaluated. Temporal recursion terminates early at position 4, as the next recursion would breach a boundary condition. In this example, $v>20$ only after the temporal recursion terminates, so $E^*_S(\Diamond_{[5,10]}\mu_{v,20},0)$ is false.}
    \label{fig:temprec}
\end{figure}

The full definitions of $E^*$ for the temporal constraints, including the boundary checks, are given below:
\[
    E^*_S(\Diamond_{[x,y]} \rho, n)  \Coloneq
    \begin{cases}
        x \leq 0 \wedge E^*_S(\rho, n)&\text{if }n= \lvert S \rvert - 1 \;\vee \\&\:\:\: y-\Delta t_n<0\\
        \\[-2ex]
        \bigl(x \leq 0 \wedge E^*_S(\rho, n)\bigr) \; \vee &\text{if }n<\lvert S \rvert - 1 \\
        \:\:\:\: E^*_S(\Diamond_{[x-\Delta t_n,y-\Delta t_n]}\rho,n+1)
    \end{cases}
\]
\begin{align*}
        E^*_S(\square_{[x,y]} \rho, n) \Coloneq &
    \begin{cases}
        x \leq 0 \wedge E^*_S(\rho, n)&\text{if }n= \lvert S \rvert - 1 \; \vee\\\phantom{ \:\:\: E^*_S(\rho_1\mathcal{U}_{[x-\Delta t_n,y-\Delta t_n]} \rho_2,n+1)\Bigr) \; \vee}&\:\:\: y-\Delta t_n<0\\
        \\[-2ex]
        \bigl(x > 0 \vee E^*_S(\rho, n)\bigr) \; \wedge &\text{if }n<\lvert S \rvert - 1 \\
        \:\:\: E^*_S(\square_{[x-\Delta t_n,y-\Delta t_n]}\rho,n+1)
    \end{cases}
\\
        E^*_S(\rho_1\mathcal{U}_{[x,y]} \rho_2, n) \Coloneq &
    \begin{cases}
        x \leq 0 \wedge E^*_S(\rho_1 \wedge \rho_2, n) &\text{if }n= \lvert S \rvert - 1 \; \vee \\&\:\:\: y-\Delta t_n<0\\
        \\[-2ex]
        \Bigl(\bigl(x>0 \vee E^*_S(\rho_1, n)\bigr) \; \wedge &\text{if }n<\lvert S \rvert - 1\\ \:\:\: E^*_S(\rho_1\mathcal{U}_{[x-\Delta t_n,y-\Delta t_n]} \rho_2,n+1)\Bigr) \; \vee \\ \:\:\: 
        \bigl(x \leq 0 \wedge E^*_S(\rho_1 \wedge \rho_2, n)\bigr) 
    \end{cases}
\end{align*}

\noindent Each time that $E^*(\rho, n)$ recurses, it either reduces the size of the STL constraint or approaches the signal termination by a single step. Isabelle proves that the lexicographic measure $\langle\lvert S \rvert - n, \lvert \rho \rvert\rangle$ strictly decreases with every recursion (where $\lvert \rho \rvert$ is the number of sub-formulae in $\rho$). Thus, $E^*$ runs in $\mathcal{O}(\lvert S \rvert \lvert \rho \rvert)$ time. This is also true of $R^*$ and $dR^*$, described in Section \ref{ssFormSTLsmooth}.

The $E^*$ function is formally specified in the Isabelle theorem prover. It is equivalent to the standard semantics for STL given in Section \ref{ssBackSTL}. The $E^*$ function thus allows us to algorithmically evaluate constraints with STL's standard semantics over a signal by recursion.

\subsection{Recursive algorithm for smooth robustness semantics of STL}
\label{ssFormSTLsmooth}

In an analogous way to how we implement the evaluation of the standard semantics, we now address formalising a robustness function. Recall that this is a function describing to what degree a signal does or does not satisfy a constraint. This makes it useful as a loss function in learning to satisfy the constraint, as it can be treated as an error.
In order to enable learning by gradient descent, this robustness function should also be differentiable.

Neither the minimum nor the maximum function used extensively in the usual robustness function (as described in Section \ref{ssBackSTL}) are differentiable everywhere. Because of this, we implement smooth, binary versions of these functions, $\max_\gamma$ and $\min_\gamma$, which use $\gamma$ as a smoothing parameter \cite{cuturi2017soft}. Their definitions are as follows:
\begin{align*}
    \max\nolimits_\gamma (a,b) &= \begin{cases}
        \max \{a,b\} & \text{if } \gamma\leq0\\
        \gamma \ln (e^{a/\gamma}+e^{b/\gamma}) & \text{if } \gamma>0\\
    \end{cases}\\
    \min\nolimits_\gamma (a,b) &= \begin{cases}
        \min \{a,b\} & \text{if } \gamma\leq0\\
        -\max_\gamma(-a,-b) & \text{if } \gamma > 0\\
    \end{cases}
\end{align*}

\noindent We prove in Isabelle that $\lim_{\gamma \to 0} (\max_\gamma (a,b)) = \max\{a,b\}$, and likewise for $\min_\gamma$. Importantly, for $\gamma>0$ both functions are differentiable everywhere with respect to either parameter $a$ or $b$.

We now specify the differentiable function $R^*_{\gamma, S}(\rho, n)$, used to compute the robustness of $\rho$ recursively over a signal $S$ from temporal position $n$, with a smoothness parameter $\gamma$ as used above. This function is equivalently referred to as a \emph{smooth semantics} for STL.

We derive the $R^*$ function from $E^*$ following similar guidelines as those used in deriving the standard robustness function $R$ from the standard semantics $E$ (discussed in Section~\ref{ssBackSTL}). We replace disjunctions in $E^*$ with $\max_\gamma$ in $R^*$, and conjunctions with $\min_\gamma$. Where $E^*$ checks if $x\leq0$ we simply use $-x$ (and $x$ if checking $x>0$). We give an example of this translation below using the Always constraint:
\[
R^*_{\gamma,S}(\square_{[x,y]}\rho, n) =
\begin{cases}
    \min_\gamma\bigl(-x,R^*_{\gamma,S}(\rho, n)\bigr)
    & \text{if } n=\lvert S \rvert -1 \; \vee \\
    & \:\:\: y-\Delta t_n<0\\
    \\[-2ex]
    \min_\gamma\Bigl(
        \max_\gamma\bigl(
            x,
            R^*_{\gamma,S}(\rho, n)
        \bigr),& \text{if } n<\lvert S \rvert - 1\\\:\:\:
        \qquad{}\;R^*_{\gamma,S}(\square_{[x-\Delta t_n,y-\Delta t_n]}\rho, n+1)
    \Bigr)
    
\end{cases}
\]

\noindent We now have a function that gives us an algorithm to compute robustness. As already mentioned, robustness can tell us how well a constraint is satisfied, or how badly it is unsatisfied. This second use can be adapted as a loss function. But we need to go further to meet the needs of learning that uses gradient descent, by having an algorithm to compute the derivative of robustness with respect to any state variable of the signal.

We proceed to build this derivative function $dR^*_{\gamma,S}(\rho,n,v_{i,k})$, which calculates the derivative with respect to $v_{i,k}$, the $i^{\text{th}}$ state variable at temporal position $k$. We derive $dR^*$ using the chain derivative rule and the derivatives for $\max_\gamma$, $\min_\gamma$ and the atomic constraint. The derivatives for $\max_\gamma$ (and $\min_\gamma$) take four parameters: the values being compared and their derivatives. For example, the function $\max_\gamma(a, b)$ has a derivative calculated using the function $\text{dmax}_\gamma(a, da, b, db, v_{i,k})$, where $da$ and $db$ are the derivatives (with respect to $v_{i,k}$) for the functions used to calculate $a$ and $b$. We formally prove with Isabelle that this derivative is correct under the assumption that $da$ and $db$ are correct.

We present an illustration of how $dR^*$ is defined using the Always constraint as an example (compare with the $R^*$ definition above):

\[
dR^*_{\gamma,S}(\square_{[x,y]}\rho, n, v_{i,k}) =
\begin{cases}
    \text{dmin}_\gamma\bigl(-x,0,
    & \text{if } n=\lvert S \rvert -1 \; \vee \\
    \:\:\:R^*(\rho,n), dR^*(\rho, n, v_{i,k})\bigr) & \:\:\:y-\Delta t_n<0\\
    \\[-2ex]
    \text{dmin}_\gamma\Bigl(
        \max_\gamma\bigl(
            x,
            R^*_{\gamma,S}(\rho, n)
        \bigr),& \text{if } n<\lvert S \rvert - 1\\\:\:\:
        \text{dmax}_\gamma\bigl(x,0,R^*_{\gamma,S}(\rho,n),dR^*_{\gamma,S}(\rho,n,v_{i,k}),v_{i,k}\bigr),\\\:\:\:
        R^*_{\gamma,S}(\square_{[x-\Delta t_n,y-\Delta t_n]}\rho, n+1),\\\:\:\:
        dR^*_{\gamma,S}(\square_{[x-\Delta t_n,y-\Delta t_n]}\rho, n+1,v_{i,k}),v_{i,k}
    \Bigr)
    
\end{cases}
\]

\noindent We can then use $dR^*$ with any gradient descent method to learn how that variable can be changed to maximise robustness.

As with the $E^*$ function, both the $R^*$ function and its derivative $dR^*$ are formally specified in the Isabelle theorem prover. We use Isabelle to formally prove two important theorems about these functions, namely the soundness and the correctness of the derivative:

\begin{theorem}[Soundness] $$\lim_{\gamma\to0}R^*_{\gamma,S}(\rho,n)>0 \implies E^*_S(\rho, n)$$
$$\lim_{\gamma\to0}R^*_{\gamma,S}(\rho,n)<0 \implies \neg E^*_S(\rho, n)$$
\end{theorem}

\begin{theorem}[Derivative correctness] Assuming $\gamma>0$,
$$\frac{d\bigl(R^*_{\gamma,S}(\rho,n)\bigr)}{dv_{i,n}}=dR^*_{\gamma,S}(\rho,n,v_{i,n})$$
\end{theorem}

The proofs of both these theorems proceed by induction on the size of the constraint and the size of the signal being examined.
In the base case we establish the conclusion for the simplest constraint at a single time index.
In the inductive step we extend the conclusion from smaller constraints and shorter segments of the signal to increasingly larger constraints and segments.
Both of these directions of induction are vital to cover both the propositional and temporal aspects of the logic.
Initially, we establish the soundness without smoothing (i.e.\ $\gamma = 0$) and expand this result to $\gamma>0$ by using continuity of $R^*$.

The full formal proofs of these facts, which are part of an Isabelle formalisation that is too extensive to cover in this article, provide us with strong, \emph{a priori} guarantees about our overall approach.

\subsection{Code generation}
\label{ssFormCodeGen}

We use the code generation capabilities of Isabelle to export our specifications as working OCaml code \cite{haftmann_codegen}. This means that the structure of our formal definitions is translated via a thin layer of trusted equivalences into OCaml code, without additional human intervention.
This translation layer is small and comes with partial correctness proofs \cite{10.1007/978-3-642-12251-4_9}, and so we have strong confidence that the proofs we have established for our Isabelle specifications remain valid for the code that is generated.

As we discussed in Section \ref{ssBackSTL}, our definition of STL is not minimal. We include the Eventually and Always constraints as primitives instead of defining them via the Until constraint. This means our generated code is more efficient, as our approach prevents a complicated chain of constraints slowing down calculations during code execution.

The code is then called from an instance of the \texttt{torch.autograd.Function} PyTorch class, with $R^*$ as its forward method, and $dR^*$ as its backwards method \cite{chevallier2025formallyverifiedneurosymbolictrajectory,paszke2019pytorch}. This can then be handled like any other function used by PyTorch for optimisation via gradient descent, and incorporated into neurosymbolic learning.

\section{Case Study}
\label{sExperiments}

We replicated many of the experiments from the STLCG paper \cite{leung2023backpropagation}, then extended these with one that uses an irregularly sampled signal (which is out of scope for STLCG). In the latter, described below, we demonstrate how an agent can learn to respect pre-specified (spatio-)temporal rules in a complicated and irregularly sampled environment. 

Consider this setting: a medical robot, assisting a disabled patient in a single room, which must perform certain tasks in a 50 second time limit while also avoiding risks to the patient and to the structure of the room. All tasks begin with the robot in a docking position. The specific task we use in our demonstration requires that the robot go from the docking position to a medical cabinet, retrieve some medicine, and then administer it to the patient in bed, before returning to its dock. To complete the retrieval and administration tasks the robot must remain in the correct area for 5 seconds (we do not specify the exact requirements of these fine motor control tasks as we are only concerned with navigation in our problem). At all times, the robot must be careful to avoid furniture (the chair, a desk, and the bed), and not to exceed the speed limit $v_m$.

We represent the task by the following STL constraint, annotated with an informal interpretation for each of the clauses that need to be simultaneously satisfied: \begin{flalign*}
\square_{[0.0,50.0]}\mu_{-v,-v_m}\ \wedge &&\text{The whole time follow the speed limit, }v<v_m\text{ and}\\
\square_{[0.0,50.0]}\neg\textbf{Desk}\ \wedge &&\text{avoid the desk area for the whole time period, and}\\
\square_{[0.0,50.0]}\neg\textbf{Chair}\ \wedge &&\text{likewise avoid the chair area, and}\\
\square_{[0.0,50.0]}\neg\textbf{Bed}\ \wedge &&\text{likewise avoid the bed area, and}\\
\Diamond_{[0.0,50.0]}\Bigl(\square_{[0.0,5.0]}\textbf{Access}\ \wedge &&\text{access the cabinet for 5s to gather medicine, and}\\
\Diamond_{[0.0,50.0]}\bigl(\square_{[0.0,5.0]}\textbf{Bedside}\ \wedge &&\text{\emph{then} go to bedside for 5s to administer medicine, and}\\
\Diamond_{[0.0,50.0]}(\square_{[0.0,50.0]}\textbf{Dock})\bigr)\Bigr) &&\text{\emph{then} return to the dock for any remaining time.}
\end{flalign*}

\noindent For our speed limit, recall that at any sample $n$ with velocity $v_n$, $E^*(\mu_{-v,-v_m},n) \iff -v_n>-v_m \iff v_n < v_m$. Each constraint in boldface above is an abbreviation for a conjunction of atomic constraints on the $x,y$ co-ordinates, which are satisfied if the robot is within a specific region (either rectangular or circular).

Planning the trajectory to finish each task involves plotting a path through the room. We can sample from this path to produce a signal, but this sampling need not be uniform. Physical constraints (disruption to transmission, for example) may introduce randomness to the sampling times. Additionally we may wish to sample more often when there is a higher risk of breaching a constraint, and less often when we are in safer areas.

The trajectory to learn is initialised as a set of straight lines going from each task location to the next, ignoring possible risks and time-requirements for tasks, and assuming uniform travel speed. The \emph{initial signal} is found from this path by sampling fifty times, each time taking two state variables directly from the path, the $(x, y)$ co-ordinates. A third state variable is computed from these and the time indices of the samples, representing the speed $v$. We introduce non-uniform sampling by sampling more frequently where there is higher risk of constraint breach -- in particular in the second section of the path between the cabinet and the bed. We know the fixed room layout, so we can determine where this risk is highest by seeing where the initial path crosses through obstacles most frequently.

\begin{figure*}[hbtp]
    \centering
    \includegraphics[width=0.5\textwidth]{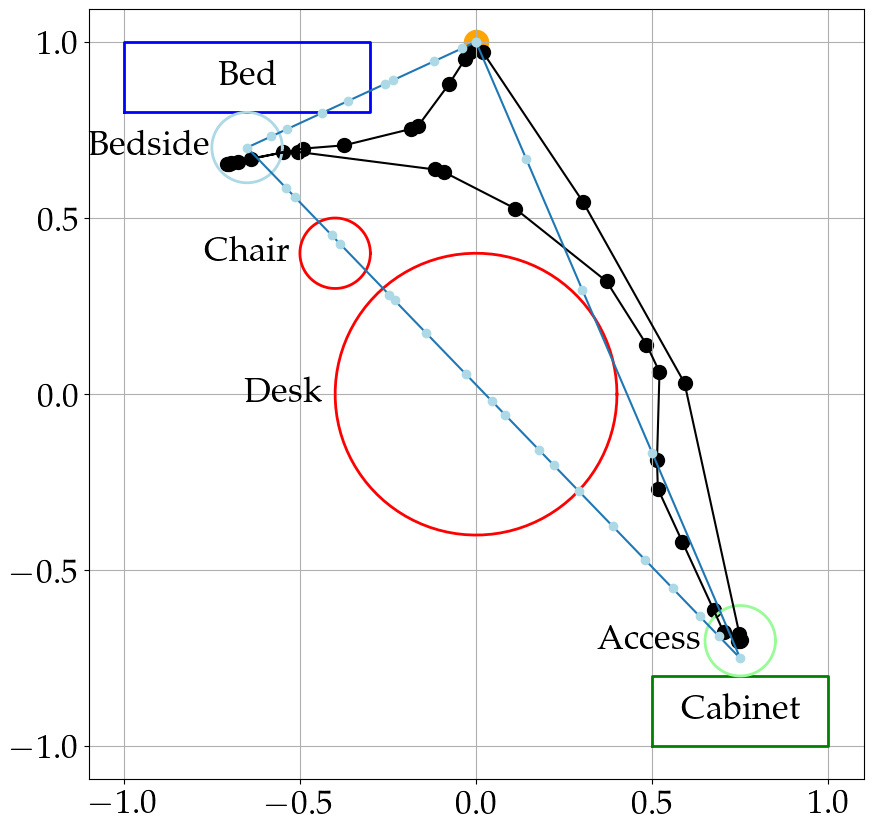}
    \caption{Result showing the agent accessing specific points and navigating the obstacles in the environment. The light blue line is the unaltered original path, with dots indicating when samples were taken for the initial signal. The black line and dots are the trained final path and  signal.}
    \label{fig:results}
\end{figure*}

The STL constraint we are learning from is not simple: it involves extracting different kinds of data in a path, and nesting temporal constraints. In addition, the signal is sampled irregularly and more frequently where we suspect the path has higher risk, as discussed above. Nonetheless, as GradSTL is formally verified, we have confidence that the signal will be correctly evaluated by it, and this result is what we see in Figure \ref{fig:results}.

GradSTL computes the robustness of the signal and its derivative for each state variable in the signal, using the STL constraint above. This is then optimised using the Adam optimiser to adjust the signal \cite{adam}. This process is repeated 500 times. The difference between the initial and the result signals, illustrated in Figure \ref{fig:results}, clearly demonstrates that the optimisation process has learned to satisfy the constraint.

\section{Conclusion}
\label{sConclusion}

To our knowledge, GradSTL is the first implementation of STL suitable for neurosymbolic learning over any signal. Our formally verified approach using theorem proving gives great confidence that it is both comprehensive and sound with respect to the standard semantics of STL. Experimental work in our case study demonstrates that it enables successful learning even with complicated constraints and irregular sampling of the signal.

GradSTL overcomes the limitations of previous work, namely implementations of STL which restrict the class of supported constraints or signals.
This is possible because our adaptive temporal window technique adjusts STL constraints during temporal recursion. This allows us to identify the correct times that need to be evaluated by any temporal constraint, regardless of when it is first evaluated and whether it is nested.

GradSTL uses a formally specified, recursive algorithm for smooth STL semantics to automatically generate faithful code. Neurosymbolic AI is a growing research area, and finding ways to directly integrate logic and domain knowledge into learning will be of increasing importance. GradSTL is a step in that direction.

\subsection*{Acknowledgements}

The final version of this paper will appear in the proceedings of the 32nd International Symposium on Temporal Representation and Reasoning (TIME 2025). We thank the TIME 2025 referees for their feedback. \medskip

\noindent This research was funded by the Edinburgh Laboratory for Integrated Artificial Intelligence (ELIAI) EPSRC (EP/W002876/1), by the Fonds National de la Recherche, Luxembourg (AFR 15671644), and by the Legal \& General Group (research grant to establish the independent Advanced Care Research Centre at University of Edinburgh). Legal \& General had no role in conducting the study, interpretation or the decision to submit for publication. The views expressed are those of the authors and not necessarily those of Legal \& General.

\bibliography{constrainedtraining}
\bibliographystyle{theapa}

\end{document}